# WebParF:A Web Partitioning Framework for Parallel Crawler

Sonali Gupta

Department of Computer Engineering
YMCA University of Science & Technology
Faridabad, Haryana 121005, India
Sonali.goyal@yahoo.com

Komal Bhatia

Department of Computer Engineering
YMCA University of Science & Technology
Faridabad, Haryana 121005, India
Komal_bhatia11@rediffmail.com

Pikakshi Manchanda

Student, M.Tech, Department of Computer Engineering
YMCA University of Science & Technology
Faridabad, Haryana 121005, India
Pikakshi787@gmail.com

*Abstract*—**With the ever proliferating size and scale of the WWW [1], efficient ways of exploring content are of increasing importance. How can we efficiently retrieve information from it through crawling? And in this "era of tera" and multi-core processors, we ought to think of multi-threaded processes as a serving solution. So, even better how can we improve the crawling performance by using parallel crawlers that work independently? The paper devotes to the fundamental development in the field of parallel crawlers [4], highlighting the advantages and challenges arising from its design. The paper also focuses on the aspect of URL distribution among the various parallel crawling processes or threads and ordering the URLs within each distributed set of URLs. How to distribute URLs from the URL frontier to the various concurrently executing crawling process threads is an orthogonal problem. The paper provides a solution to the problem by designing a framework WebParF that partitions the URL frontier into a several URL queues while considering the various design issues.**

## I. INTRODUCTION

The Web is larger than it looks, with millions and billions of web pages that offer informational content spanning hundreds of domains and many languages across the world. Due to the colossal size of the WWW [1], search engines have become the imperative tool to search and retrieve information from it [2]. The typical design of search engines is a "cascade", in which a Web crawler creates a collection which is indexed and searched. Most of the designs of search engines consider the Web crawler as just a first stage in Web search, with little feedback from the ranking algorithms to the crawling process. This is a cascade model, in which operations are executed in strict order: first crawling, then indexing, and then searching. A basic model of a typical search engine is given in figure 1.

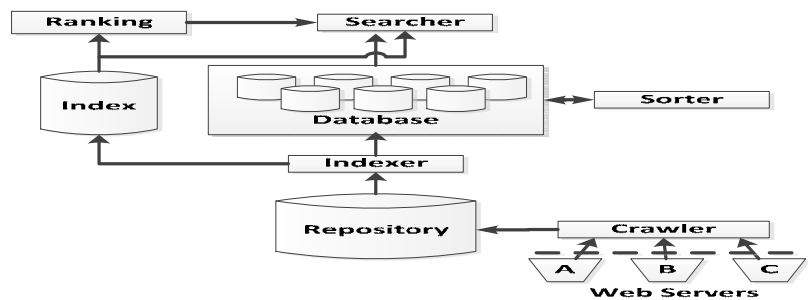

Figure 1.   General Search Engine

The web crawler is an automatic web object retrieval system that exploits the web's dense link structure. It has two primary goals, to seek out new web objects, and to observe changes in previously-discovered web objects (web event detection).The basic web crawler algorithm has not changed since the World Wide Web Wanderer (the first reported web crawler) was designed in 1993. Almost all crawlers follow some variant of the basic web-traversal algorithm. The working principle behind any basic crawler can be represented with the below figure 2.

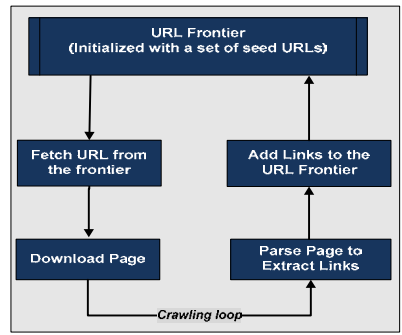

Figure 2.    General Crawler

It starts by placing an initial set of URLs called as seeds in a queue called the URL frontier and then picking a URL from this seed set, fetch the web page at that URL and finally parse it to extract the text and hyperlinks to other web pages or objects. The extracted links are then added back to the URL frontier which comprises of URLs whose corresponding web pages have still to be fetched by the crawler. The process is repeated by the crawler until it decides to stop which may either be due to no more URLs left for downloading in the URL frontier or achieving the target number of downloaded documents or some other resource limitations.

As the size of information on the Web flourishes, it becomes not only difficult but also impossible to crawl the entire WWW; visiting this huge, distributed information source under the specified time bounds demand efficient and smart crawling techniques, as sequential crawling will not suffice for more than just a small set of the entire WWW contents. Alternatively, it can be said that a single crawling process cannot scale to such massive amount of data on the WWW [4, 5, 14].

The paper is organized as follows: Section 2 throws light into the design issues related to any parallel crawler, Section 3 focuses on the basic design of our proposed framework for URL distribution which is an essential substrate for any of the parallel crawling tasks, Section 4 provides a software architecture for the same and finally Section 5 concludes the paper with a word to the future work.

## II.    PARALLEL CRAWLERS

To address the shortcoming of the single sequential crawler, multiple crawling processes (referred to as C-procs) are employed in parallel, where each such process loads a distinct page in tandem with others and performs all the basic tasks that are performed by the single process crawler as mentioned above in section 1. Figure 3 depicts a basic architecture for parallel crawler [4]. Each parallel crawler has its own database of collected pages and own queue of un-visited URLs. Once the crawling process is over, the collected pages of all the C-Procs are added to the search engines index. More such additional crawling processes, C-procs can be put in the crawling system if the desired download rate has not been achieved and needs to be increased further.

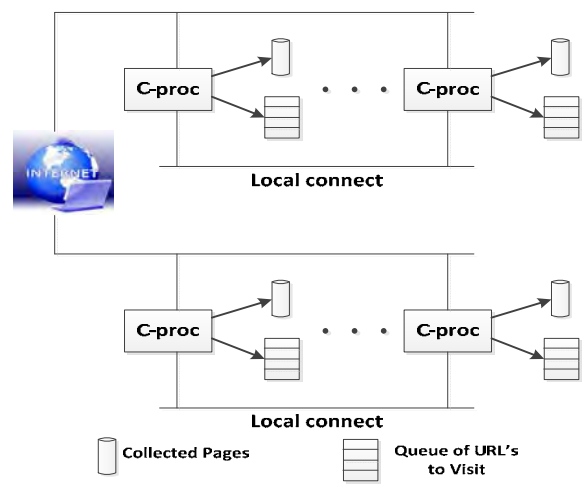

Figure 3.    Generic Architecture of a Parallel Crawler

The potential of a parallel crawler in terms of throughput and quality is immense; we envision the following advantages of such a crawler:

- **Scalable:** Due to enormous size of the Web, it is often imperative to run a parallel crawler. A single-process crawler simply cannot achieve the required download rate in certain cases.

- **Network-load dispersion:** Multiple crawling processes of a parallel crawler may run at geographically distant locations, each downloading "geographically-adjacent" pages. For example, a process in Germany may download all European pages, while another in Japan crawls all Asian pages. In this way, we can disperse the network load to multiple regions. In particular, this dispersion might be necessary when a single network cannot handle the heavy load from a large-scale crawl.

- **Robust:** The information is published and made available on the web in an uncontrolled way. A parallel crawler is robust enough to handle the hazardous situations that may affect its performance or cause mal-functioning, by putting on additional processes.

Many search engines have implemented their own versions of parallel crawlers to index the Web like: The author in [12] presents Mercator which is a scalable and extensible crawler, rolled into the AltaVista search engine; The Ubicrawler as described in [14] is a high performance fully distributed crawler in terms of the crawling processes with a prime focus on distribution and fault tolerance. The basic design criteria involved behind creating all these crawlers remained the same along with a common set challenges faced. However, still remains very little scientific research on the topic. The next section presents the various challenges to be addressed for an effective and efficient design of such crawlers and the aspect from which they have been looked at to attain a possible solution.

### III. Problem Statement

Creating a favorable architecture for parallel crawlers requires addressing many challenges. Some of the basic but the most important of those challenges are:

- **URL duplication:** The problem occurs when list of URLs to be retrieved by the crawling system. The list will be overseen by all the parallel processes; it can be ensured that the same URL is not retrieved multiple times.

- **Content duplication:** When the same web page, referred to by different URLs is retrieved by multiple crawling processes. Content duplication can be reduced by maintaining the retrieved data at a centralized server. However, there may be a situation when the central server itself acts as a single source of failure.

- **Communication Overhead:** In order to maintain the quality of the downloaded collection of web pages, the overlap in URLs and content by the different C-procs should be minimized. This minimization is possible only when individual crawling processes communicate among themselves to coordinate the whole crawling process. This communication yields an inevitable overhead by consuming not only crawler's time but also the network bandwidth.

A strategy that makes satisfactory trade-off among these objectives is required for an optimized performance. All the above challenges can be gathered and named as "the problem of creating web partitions". In other terms, the problem can be stated as:

Finding a solution to the Web Partitioning problem will help us to conquer all these questions. In the section that follows, we present the architecture of our system for Web partitioning, inclined towards the dynamic and distributed nature of the WWW.

### IV. System Design & Architecture

The overlap as already mentioned, can occur either in URL or in content, forming the basis for the two types of web partitioning schemes [11, 15, 16]: URL-oriented partitioning and content oriented partitioning. The methods differ in the order in which the partition assignment is determined with respect to the retrieval of web pages.

- **URL-oriented partitioning:** The scheme divides the web by assigning URLs to partitions. They eliminate URL duplication as the same URL is always retrieved by the same crawler process and never by any other crawling process. The kind of between the crawler process threads is the URL. The URL-oriented partitioning would result in multiple crawlers obtaining copies of identical web pages for the reason that a web page can be referred to by different URLs.

- **Content oriented partitioning:** This scheme assigns a partition after A has been retrieved when the crawler is provided with the URL. The retrieved web page is then transforming the multi-topic web page into several single topic small units which are then transferred to the corresponding crawler processes.

Our partitioning approach is a combination of both the schemes so as to eliminate the overlap both in URL and in content. The adopted approach is influenced by the fact that web pages are significantly more likely to link to pages that are related to the domain of the containing web page, as opposed to pages that are selected at random [3,7,8,10]. Also, if a crawling process is designated to retrieve pages from a specific domain, it is likely to continue retrieving pages from that domain [8]. Nonetheless, the prime reason for creating these Web partitions lies in facilitating the development of an efficient parallel crawler, thus WebParF divides the web into various domains of general interest, based on the idea of domain specific crawling that ascertains On downloading the webpage corresponding to the assigned URL, the crawling process retains only the resources that belong to its domain of concern while routing the rest to the appropriate crawling processes. So, whenever a crawling process retrieves a web page, it has to examine the page to determine its domain and predict its relevancy [11] to one of the many domains of the various parallel crawling processes. This will help us to reduce the redundancy in content of the crawler as a crawling process will not try to download a page that does not belong to its domain. Thus our approach is a domain based or domain oriented partitioning approach that is likely to make the system highly scalable as the crawling process threads may be added by breaking a domain into sub-domains.

To create domain specific partitions of the URL frontier our system has been divided into two phases, the former of which involves partitioning the seed URL frontier that gives an illusion of independent URL frontiers to each of the various concurrently executing threads of the parallel crawler while the latter simply stresses on allocating the URLs from the independent seeming URL frontiers to the various crawler threads. Figure 4 provides the architecture of our system, WebParF.

The main aim of our partitioning system is to facilitate the design of a parallel crawler where the various crawling processes have been dedicated to the common task of "concurrently crawling the Web" while forcing them to work as independently as possible. The independent working of each parallel crawling process can only be achieved if they are assigned a unique identity in themselves. This has been done in our system by dividing the URL frontier that gives an illusion of an independent or dedicated frontier to each of the crawling processes of the parallel crawler. Thus, each crawling process will be assigned its portion of the Web to crawl.

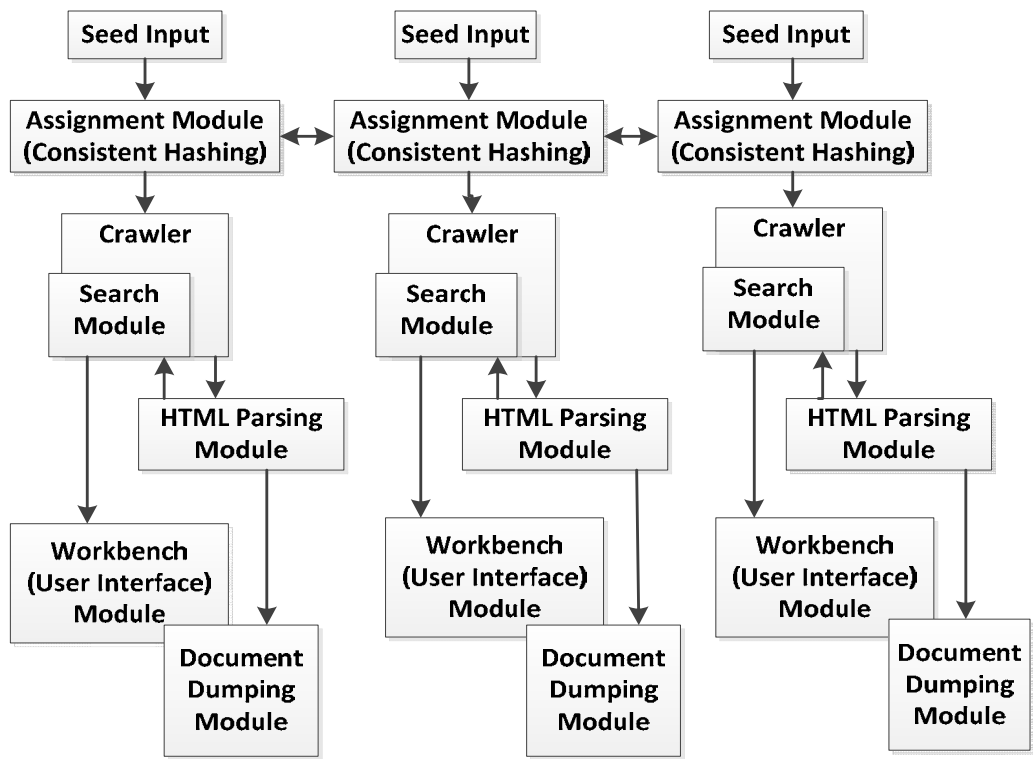

Figure 4. Architecture of WebParF system

A. Phase I: Creating a URL frontier for parallel access

This phase is the most basic and important to the entire process and is used to set the seed URLs of the URL frontier. Due to the concurrent access of the URL frontier by a number of parallel crawling processes, it will be known as the Global URL Frontier. The first step for the creation of the Global URL Frontier is to gather the initial seed URLs, divide the seeds into different groups one representing each domain. Each such group will here be referred as Domain Specific URL Pool. The next step is to assign priorities to URLs in each of the Domain Specific URL pool so as to develop several priority queues of URLs for the various domain specific URL pools. This sequence of steps can be described as :

*1)* Gathering the initial seed URLs for creating the Domain Specific URL pools

To gather the seed URLs for the domain specific URL Pools, two different methods are used by the interface of our proposed crawler.

The first method suggests taking advantage of the CLASSIFICATION HIERARCHY of any trusted search engine like Google since such a directory includes a collection of Websites selected and manually classified by Open Directory volunteer editors instead of constructing a new classification scheme manually. So for the first run of our crawler, the seeds are the different "hub" pages highly relevant to the various domains under consideration. A "hub" page is a page that consists primarily of links to other documents while providing little or no useful content of its own. For each category in the hierarchy the system supports the retrieval of N top relevant URLs. Those top N pages will serve as starting points for the crawler. Each domain specific URL POOL consists of some domain specific keywords and links to the Web pages designated as the starting point in a domain. The URL pool also keeps a record of the number of times a specific URL has been requested.

The other method is gathering the URLs from the previous crawls performed by the crawler. Prior to the first crawl to be performed by the crawler, seeds are only available through the first method. For the later runs, each extracted URL will be identified for its domain and then added to the respective domain specific URL pool.

*2) Ordering the URL pool*

The URLs in the URL pool need to be ordered for relevancy so as to enable the crawler to always populate its document collection with some finite number of top relevant pages in that domain [12, 13].

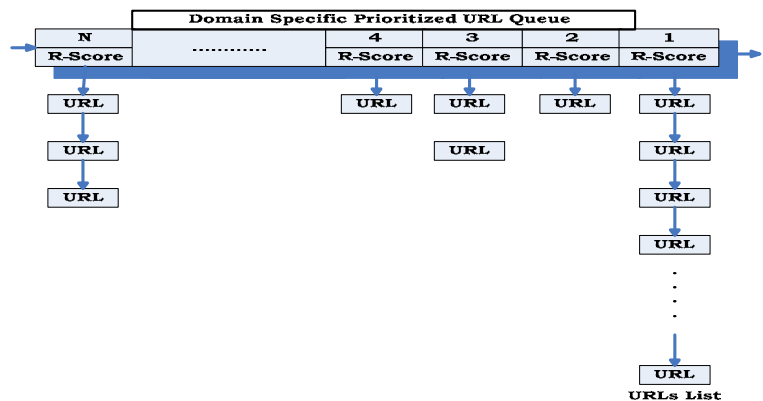

Figure 5. URL Queue

This task is accomplished with the help of a *URL ranker* module in our system, which assigns a relevance score to each URL in a specific domain pool and then places the domain identified URL in its respective URL frontier at the specified position, along with the relevance score assigned to it. The URL frontier acts as a prioritized queue for the URLs in the URL pool belonging to its domain of concern

The relevance score of a URL can be computed by using some kind of a scoring mechanism based on the metrics like the number of pages linking to the URL, how many times the URL has been requested etc. The structure of the Prioritized URL queues that has been used in our approach is as above:

Thus, the URL ranker is responsible for maintaining various prioritized URL frontier queues, one corresponding to each domain specific URL pool. The sequence of URLs in each URL Pool is sorted on the basis of the assigned relevance score .The URL that seem most relevant in the domain of search occupies the first position in the priority queue. The URLs with the same relevance score are then grouped to form a list. So, each node of the prioritized URL queue stores the position, relevance score and has a pointer to the URL list for that relevance score. This list pointed to by the pointer is accessed in the same way as a FIFO queue. The URLs at the next position in the priority queue will be the ones that are less relevant but following a hyperlink in the corresponding web page will lead to a relevant page.

As specified, all these tasks together will create a set of multiple prioritized URL seed queues, with the first one being taken from each queue, for processing in parallel with the others.

*B. Phase II: URL Allocation and Collection Creation*

This is the next phase of our system that is responsible for selecting URLs from the URL frontiers, opening the network connection to retrieve the URLs from the Web, and analyzing the downloaded web pages for finding their domains. This is accomplished with the help of following functional modules of this phase:

*1)   URL Allocator*

The URL allocator is responsible for disseminating URLs from multiple prioritized URL frontiers to the individual threads of the document loader. It dynamically assigns multiple URLs, one from each domain, to different downloader threads so as to avoid downloading duplicate web pages.

```
URL_Dispatcher()
Begin
Do forever
Begin
While (url seed queue not full)
Begin
Read url IP pair from database;
Store it into url IP queue;
End
End

URL Distributor()
While (true)
If (not empty (seed queue)) then
Src_cw= find domain (URL)
For each cw in all_crawlworker
If (not full (DNSqueue_of_src_CW)) then
Distribute(src_cw,url)
```

Figure 6.   URL Allocator Algorithm

*2)   A Multi-threaded document loader*

Since, the Web is based on a client server architecture designed for multiple, distributed clients that access a single server , so in our next step, the multithreaded downloader which is a high performance asynchronous HTTP client capable of downloading hundreds of web pages in parallel, sets off a number of downloading agents equal to the number of prioritized URL frontiers .The multiple downloading agents  start downloading pages in parallel from the multiple web servers and stores them onto the storage space provided.

```
Downloader ()
begin
do for ever
wait (Download WP);
while (empty (LOSBuffer))
begin
extract a URL;
download Webpage;
store webpage in webpage file
repository ;
end;
signal (hungry);
end
```

Figure 7.   MT Document Loader Algorthim

*3)   Web page Analyzer*

Once the various web pages have been downloaded, we need to parse their contents to extract information that will be fed to possibly guide the future path of the crawler. So, each such downloaded Web page is given as input to one of the various threads of the Web page Analyzer, each of which consists of two components: a parser and a page classifier.

The role of the parser is to extract the various HTML components of the page. For each such downloaded page, the parser parses it for various HTML tags, scripts and text. The basic job of any parser is to extract list of new URLs from the fetched or downloaded Web page and return the new-unvisited URLs to the URL frontier. The Link address or the URL of other Web resources is specified in the href attribute of the anchor tag (indicated by <a>) similar to the one shown here:

**<A HREF= "http://www.w3schools.com"> this is a Link</A>**

So, the parser must extract the anchor tags that supports href attribute for future crawling. All the extracted URLs are stored temporarily in the URL database for further use by the crawler.

The Classifier component identifies the domain of the web page and adds the web page to the associated repository of its domain. The domain of a web page can be found by using the technique described in our earlier work in [9]. It also tags the URL of the just downloaded web page with the achieved domain and adds this information into the URL database, the same database that is used to store newly discovered URLs. Tagging the URL with the domain information of its web page will prove beneficial in identifying the domain of any new URL during the later step of adding the discovered URLs to the domain specific URL pool.

*4) URL Dispatcher*

The URLs in the Web page detected by the Web Page Analyzer must be added to the Global URL Frontier. But instead of directly adding to the Global URL Frontier they have been temporarily stored in a URL database where they undergo processing by the URL dispatcher component for finding and adding more details about a URL. It serves the following purposes:

- The component first restores relative hyperlinks to absolute hyperlinks. Since, the same URL may be referenced in many different documents restoring any relative hyperlinks to their absolute value is important.

- As a next step, it tries to predict the domain of the discovered URLs. This has somewhat been made easier by the tagging of URLs as these tagged URLs might be serving as the source of certain discovered URLs. This can be assumed from the hyperlinked structure of the Web and the fact that web pages are significantly more likely to link to pages that are related to the domain of the containing web page. Predicting the domain of the URL can also avail the benefit from any information that seems useful and has been discovered during the current or any of the previous crawls of the crawler.

- It then filters the new URLs against any duplicacy with the domain specific URL pool. Finally if no duplicacy is exhibited, the dispatcher adds the new URL to its corresponding domain specific URL pool in the Global URL Frontier. Now at this point exact duplicates will be eliminated by the URL dispatcher component to preserve network bandwidth. The overlap is avoided at any early stage when crawler executes next, with the help of this URL Dispatcher component.

Also, during crawling it is not necessary to add the newly found URLs to the Global URL frontier each and every time a Web page is parsed. These new URLs are usually added periodically or in batches because:

- ***Index of a search engine is created & updated in batches***: The goal of the crawling process is to create a collection of pages that is useful for the search engine's index. So, the fraction of the web that a crawler downloads should represent the most important pages to be indexed. This process of indexing is done in batches as it is very inefficient to do it one web page at a time and if not done so the index construction becomes a bottleneck. Thus, in most search engines, the index is not updated continuously, but rather updated completely at some later time [5]. To the best of our knowledge, this is the case for most large search engines like GOOGLE. When the index is updated in batches, it is not important which URLs were transferred first.

- ***Download the important URLs earlier in the crawl:*** "Good" pages are seen early in the crawling process [NW01]. Conversely, if a URL is seen for the first time in a late stage of the crawling process, there is a high probability that it is not a very interesting page. Adding it alone to the Index hardly makes any change to the value of the Index. An altogether updation in the index may prove to be valuable.

- ***URLs Exchanged in groups for certain crawlers:*** If the crawler is distributed or centrally coordinated by a server, then it has to send the results back to the central server, or exchange results with other crawling processes. For better efficiency and performance, many URLs should be sent or exchanges simultaneously. This will reduce the context switch overhead and thus lead to better performance.

The periodic addition of newly found URLs to the Global Frontier creates the need for a temporary storage area that will store the discovered URLs for the meantime which in our architecture has been served by the URL database.

## CONCLUSIONS

The vast amount of information that is available online has urged to develop efficient and domain specific crawlers [3, 7] that can download a large fraction of the Web in a desired small unit of time. This necessitates an approach that is not only efficient but also highly scalable. The paper proposed a novel approach that efficiently navigates through the URLs by organizing them according to their suitable domains and partitioning the URL frontier into multiple prioritized URL queues so that the URLs are distributed among the parallel crawling processes to eliminate overlap in content. Our partitioning system facilitates the design of a parallel crawler by making it not only scalable through division by domains and sub-domains, but also fault tolerant by making a balanced distribution of load among all the remaining crawler process threads that were held responsible for harvesting the pages from the same domain as that of the dying process. The system can be improved further if the domain information of all the URLs can be made available prior to fetching the associated page from the Web. . The system is further expected to improve if a better design of the classifier and the dispatcher modules is framed A possible solution towards this end can be achieved through the extraction of all the back-links and for-links of the candidate URLs to be fetched.

## REFERENCES

[1]  Lawrence,S.; Giles, C.,L.: Searching the World Wide Web. Science, Vol.280, pp. 98-100, (1998) www.sciencemag.org
[2]  Baeza-Yates, R., & Ribeiro-Neto, B. (1999). Modern information retrieval (2nd ed.). Addison-Wesley-Longman
[3]  Chakrabarti,S.; Berg, M. van den; Dom,B.: Focused crawling: a new approach to topic-specific web resource discovery.Computer Networks, 31(11-16) ,p.p.1623–1640, 1999.
[4]  J.Cho, H.Garcia Molina.. Parallel Crawlers. In Proceedings of the ACM Annual Conference WWW , 2002 Hawaii USA
[5]  Brin, S., Page, L., 1998. The anatomy of a large-scale hypertextual Web search engine. In Computer Networks and ISDN Systems,Vol. 30, No.1-7, 107- 117.
[6]  Shkapenyuk V. and Suel T. Design and Implementation of a high-performance distributed web crawler. In Proc. 18th Int. Conf. on Data Engineering, 2002, pp. 357–368.
[7]  Arguello, J.;Diaz, F.; Callan, J.; Crespo, J.,F.:. Sources of evidence for vertical selection. In:  32nd International conference on Research and  development in Information Retrieval, SIGIR'09 pp. 315--322,  ACM, New York, USA (2009)
[8]  Qi,X.;Davison, B.D.: Knowing a web page by the company it keeps. In International conference on Information and knowledge management (CIKM), pages 228-237, 2006.
[9]  Gupta, S.;Bhatia,K.K.: A ayatem's approach towards domain identification of web pages in Second International Congerence on Parallel , Distributed and Grid Computing 2012, Page(s): 870-875 doi: 10.1109/PDGC.2012.6449938
[10] Sharma,A. K.; Bhatia, K.K.:  A Framework for Domain-Specific Interface Mapper (DSIM), International Journal of Computer Science and Network Security, VOL.8 No.12, December 2008.
[11] Marin, M.; Paredes, R.;Bonacic, C.; High Performance Priority Queues for Parllel crawlers in proceedings of the ACM conference WIDM '08 , California, USA
[12] Heydon A. and Najork M. Mercator: a scalable, extensible web crawler. World Wide Web, 2(4):219–229, December 1999
[13] Junghoo Cho, Hector Garc´ia-Molina, and Lawrence Page. Efficient crawling through URL ordering.  Computer Networks and ISDN Systems, 30(1–7):161–172, 1998.
[14] Boldi, P.;Codenotti, Massimo, S.; Vigna, S,: UbiCrawler: A scalable fully distributed web crawler.Software Pract. Exper., 34(8):711–726, 2004.
[15] Chau, D. H., Pandit, S., Wang, S., and Faloutsos, C. 2007. Parallel crawling for online social networks. In Proceedings of the 16th international Conference on World Wide Web (Banff, Alberta, Canada, May 08 - 12, 2007). WWW '07. ACM, New York, NY, 1283-1284.
[16] Shoubin Dong, Xiaofeng Lu and Ling Zhang, A Parallel Crawling Schema Using Dynamic Partition Lecture Notes in Computer Science Volume 3036/2004, pp. 287-294

## AUTHORS PROFILE

Sonali Gupta received her B.Tech degree in Computer Science & Engineering being placed in first division with Honors, from Maharshi Dayanand University, Rohtak in 2003 and M.Tech degree in Information Technology from Guru Gobind Singh Indraprastha University, Delhi in 2008. Presently, she is working as Assistant Professor in Computer Engineering Department of YMCA University of Science & Technology, Faridabad (former YMCA Institute of Engineering.) She has a teaching experience of 11 years and is currently pursuing Ph.D. in Computer Engineering from YMCA University of Science & Technology, Faridabad. The research interest includes Information retrieval, Search engines, Web Mining.

Dr. Komal Kumar Bhatia received the B.E, M.Tech. and Ph.D. degrees in Computer Science and Engineering with Hons, from Maharshi Dayanand University in 2001, 2004 and 2009, respectively. Presently, he is working as an Associate Professor in Computer Engineering Department of YMCA University of Science & Technology, Faridabad (former YMCA Institute of Engineering) he has guided several M.Tech thesis and is also guiding Ph.Ds in Computer Engineering. His areas of interests include information retrieval, Search Engines, Crawlers and Hidden Web. He has several publications in national and international conferences. He has also contributed articles to several international journals of repute.

Pikakshi Manchanda received her Btech degree in Computer Scinece and Engineering from MDU Rohtak.Presently she is pursuing M.Tech from YMCAUST Faridabad.